\documentclass[preprint,showpacs,
preprintnumbers,amsmath,amssymb]{revtex4}

\usepackage{etex}

\usepackage{amssymb,amsthm,amscd,amsbsy,array}
\usepackage{graphicx}
\usepackage{bm}
\usepackage{amsmath,dsfont}

\usepackage{graphics,graphicx,xcolor}
\usepackage[all]{xy}

\usepackage[colorlinks=true, pdfstartview=FitV, linkcolor=blue, citecolor=blue, urlcolor=blue]{hyperref} 

\numberwithin{equation}{section}

\let\ssection=\section
\renewcommand{\section}{\setcounter{equation}{0}\ssection}

\newcommand{\Ad}{\mathrm{Ad}}

\newcommand{\barbc}{{\bc^{T}}}

\newcommand{\bb}{{\bf b}}

\newcommand{\bbeta}{\boldsymbol{\beta}}
\newcommand{\bfeta}{\boldsymbol{\eta}}

\newcommand{\bc}{{\bf c}}

\newcommand{\bbC}{\mathbb{C}}
\newcommand{\cC}{{\mathcal{C}}}
\newcommand{\BMS}{{\mathrm{BMS}}}

\newcommand{\Coad}{\mathrm{Coad}}

\newcommand{\cgal}{\mathfrak{cgal}}
\newcommand{\cga}{\mathfrak{cga}}

\newcommand{\Carr}{{\mathrm{Carr}}}
\newcommand{\carr}{{\mathfrak{carr}}}
\newcommand{\ConfCarr}{{\mathrm{CCarr}}}
\newcommand{\confcarr}{{\mathfrak{ccarr}}}
\newcommand{\hGamma}{{\widehat{\Gamma}}}
\newcommand{\pSch}{{\mathrm{Sch}}} 

\def\Barg{{{\rm Barg}}}
\newcommand{\barg}{\mathfrak{barg}}
\newcommand{\Gal}{\mathrm{Gal}}
\newcommand{\gal}{\mathfrak{gal}}

\newcommand{\opt}{\mathfrak{opt}}

\newcommand{\Diff}{{\mathrm{Diff}}}

\newcommand{\rE}{{\mathrm{E}}}

\newcommand{\veta}{\boldsymbol{\eta}}

\newcommand{\rg}{\mathrm{g}}
\newcommand{\BG}{{G}}
\newcommand{\hrg}{\widehat{\rg}}

\newcommand{\fg}{\mathfrak{g}}

\newcommand{\bgamma}{\boldsymbol{\gamma}}
\newcommand{\CGal}{{\mathrm{CGal}}}

\newcommand{\bk}{\mathbf{k}}

\newcommand{\bkappa}{\boldsymbol{\kappa}}
\newcommand{\hg}{\hat{g}}

\newcommand{\belle}{\boldsymbol{\ell}}

\newcommand{\bbN}{\mathbb{N}}
\newcommand{\cN}{{{k}}}

\newcommand{\rO}{{\mathrm{O}}}

\newcommand{\cO}{{\mathcal{O}}}

\newcommand{\bomega}{{\boldsymbol{\omega}}}

\newcommand{\CCarr}{{\mathrm{CCarr}}}

\newcommand{\ccarr}{{\mathfrak{ccarr}}}
\newcommand{\scarr}{{\mathfrak{schcarr}}}
 
\newcommand{\bp}{{\mathbf{p}}}

\newcommand{\PSL}{\mathrm{PSL}}

\newcommand{\bx}{{\bm{x}}}

\newcommand{\bbR}{\mathbb{R}}

\newcommand{\Sch}{\mathrm{Sch}}
\newcommand{\sch}{\mathfrak{sch}}

\newcommand{\SL}{\mathrm{SL}}
\newcommand{\Sl}{\mathfrak{sl}}
\newcommand{\SO}{\mathrm{SO}}

\newcommand{\se}{\mathfrak{e}}
\newcommand{\so}{\mathfrak{so}}

\newcommand{\surf}{\mathbf{surf}}
\newcommand{\sv}{\mathfrak{sv}}

\newcommand{\barbb}{{{\bb}^T}}

\newcommand{\hX}{\widehat{X}}
\newcommand{\Conf}{{\mathrm{Conf}}}
\newcommand{\conf}{{\mathrm{conf}}}

\newcommand{\Tr}{\mathrm{Tr}}

\def\smallover#1/#2{\hbox{$\textstyle\frac{#1}{#2}$}} %

\newcommand{\bu}{{\bf u}}

\newcommand{\be}{{\bf e}}
\newcommand{\bv}{{\bf v}}

\newcommand{\Vect}{\mathrm{Vect}}

\newcommand{\NU}{\mathrm{NU}}

\newcommand{\bw}{{\bf w}}

\newcommand{\bbZ}{\mathbb{Z}}

\def\bu{{\bm{u}}}
\def\bv{{\bm{v}}}

\def\beq{\begin{equation}}
\def\eeq{\end{equation}}
\def\beqa{\begin{eqnarray}}
\def\eeqa{\end{eqnarray}}
\def\nn{\nonumber}
\def\barray{\left(\begin{array}}
\def\earray{\end{array}\right)}
\def\barraynb{\begin{array}}
\def\earraynb{\end{array}}

\def\IR{{\mathds{R}}} 

\def\IC{{{C}}} 
\def\IN{{{N}}} 
\def\IS{S} 
\def\IB{{{B}}} 
\def\SL{{\rm SL}}

\def\?{\quad{\gb{\fbox{\texttt{?}}\;}}\quad}
\def\p{{\partial}}

\def\v0{\mathbf{0}}

\usepackage{color}
  
\newcommand{\blue}{\textcolor{blue}}

\newcommand{\gb}{\colorbox{green}}

\def\beq{\begin{equation}}
\def\eeq{\end{equation}}
\def\bea{\begin{eqnarray}}
\def\eea{\end{eqnarray}}

\def\p{\partial}

\def \p{{\partial}}

\newcommand{\ce}{\mathfrak{ce}}

\def\6{\partial}
\def\7{\tilde}
\def\8{\widehat}
 \def\bx{{\bf x}}

\def\G11{\Gamma_{11} }

\newcommand{\const}{\mathop{\rm const.}\nolimits}
\newcommand{\half }{\frac{1}{2}}

\begin{document}

\title{ \bf 
Conformal Carroll groups\\
}

\author{
C. Duval$^{1}$\footnote{Aix-Marseille Universit\'e, CNRS, CPT, UMR 7332, 13288 Marseille, France. Universit\'e de Toulon, CNRS, CPT, UMR 7332, 83957 La Garde, France.\\
mailto:duval@cpt.univ-mrs.fr},
G. W. Gibbons$^{2,3}$\footnote{
mailto:G.W.Gibbons@damtp.cam.ac.uk},
P. A. Horvathy$^{3}$\footnote{mailto:horvathy@lmpt.univ-tours.fr}
}

\affiliation{
$^1$Centre de Physique Th\'eorique, 
Marseille, France
\\
$^2$D.A.M.T.P., Cambridge University, U.K.
\\
$^3$Laboratoire de Math\'ematiques et de Physique
Th\'eorique,
Universit\'e de Tours,  
France
}

\date{\today}

\begin{abstract}
Conformal extensions of L\'evy-Leblond's Carroll group, based on  geometric proper\-ties analogous to those of Newton-Cartan  space-time  are proposed. The extensions are labeled by an integer $\cN$.
This framework includes and extends our recent study of the Bondi-Metzner-Sachs (BMS) and Newman-Unti (NU) groups.
The relation to Conformal Galilei groups is clarified.  Conformal Carroll symmetry is illustrated by ``Carrollian photons". Motion both in the Newton-Cartan and Carroll spaces may be related to
that of strings in the Bargmann space.\\
\bigskip
\texttt{arXiv:1403.4213 [hep-th]}
\end{abstract}

\maketitle

\baselineskip=16pt



\tableofcontents


\section{Introduction} 

In\"on\"u-Wigner contraction of the Poincar\'e group yields the
Galilei group \cite{Inonu}.
The  possibility of having another, rather unusual
contraction was pointed out  some time ago by L\'evy-Leblond, who named the result the ``Carroll group"  \cite{Leblond,SenGupta}. Despite occasional attempts \cite{Gomis,Rousseaux,Ancille,Huang}, this has long remained a sort of mathematical curiosity, though, owing to the difficulty of finding interesting physical systems which might carry such a symmetry: a massive classical mechanical system with Carrollian symmetry, for example, cannot move \cite{Gomis,Carrollvs,Ancille}.
 
The situation started to change
 when attention was drawn to the r\^ole
 of Carrollian symmetry  for tachyons \cite{thoughts,GiHaYi}.
 
Another recent line of research concerns the so-called
Conformal Galilean Algebras (CGA), initiated by a failed attempt to derive the
then new Schr\"odinger symmetry by
extending the In\"on\"u-Wigner contraction from Poincar\'e to the conformal group  \cite{Barut}, and yielding instead another ``non-relativistic conformal group'' with ``relativistic'' dynamical exponent $z=1$, now identified as ``the'' Conformal Galilean group. But this mysterious result has long been thought uninteresting as it is not a symmetry of any ``reasonable'' physical system, let alone of a free
particle. This negative outcome disqualified the subject for a long time, which has only arisen again in recent times \cite{GalConf,DHGalConf}.

Searching for physical realizations of CGA lead to general relativity \cite{Bag1} and in particular to BMS (Bondi-Metzner-Sachs) symmetry \cite{BMS,Bag2}.
 Even more recently \cite{DGHBMS}, it was recognized that the natural context of BMS symmetry is to study the conformal extensions of
 the Carroll group, and the CGA symmetry found in \cite{Bag2} is just a coincidence in one space dimension \cite{DGHBMS}. This is because the Carroll and Galilei groups in $1+1$ dimension are isomorphic, the isomorphism being effected by interchanging space and time.

In this paper is to study the various conformal extensions of the Carroll group systematically, along lines analogous to those  for CGA \cite{DHGalConf}. 

Our paper is organized as follows. To make our paper self contained and to motivate the various conformal extensions, we first summarize once again the geometric framework of Carroll group cf. \cite{Carrollvs}.

Then we proceed to a systematic discussion of various conformal extensions, insisting on the analogy between the Galilean and the Carrollian cases.

In our previous paper \cite{Carrollvs}  we constructed  massive particle models with Carroll symmetry using Souriau's method  \cite{SSD}.
The disappointing result was that the resulting dynamics is \emph{very poor}: a ``motion'' of such a particle is just one fixed point in space independently of its constant  Carrollian ``velocity''; see also  \cite{Gomis,Ancille}.

Now we show  that  \emph{massless} particle models constructed by the same method, i.e., as Carroll coadjoint orbits behave better as their motions take place on strings.
  Their ``space of motions'' is identified to the space of oriented geodesics of Euclidean $3$-space. The latter also carry a conformal Carroll symmetry with $\cN=0$.

Most interestingly, motion in both in Galilean [Newton-Cartan] and in Carroll spacetimes may be derived by considering strings in one higher dimension called  \emph{Bargmann space} \cite{DBKP,DGH}. 
Our massless Carrollian  ``photons'' correspond, in particular, to  null [also called tension-less] strings considered by Schild  in 1977 \cite{Schild}.

\section{Spacetime structures}\label{SpacetimeSection}

\subsection{Newton-Cartan manifolds}

The  geometric definition of \emph{Carroll manifolds and transformations} proposed in Refs. \cite{Carrollvs} and \cite{DGHBMS} is
 dual to that of a Newton-Cartan  manifold and its generalized symmetries \cite{GalIso,DHGalConf};  therefore we first remind  the reader how the latter are defined.

The \emph{weak definition} of Newton-Cartan (NC) manifold is that it is a triple 
$
(\IN,\gamma,\theta),
$
where $\IN$ (for Newton) is a smooth $(d+1)$-dimensional manifold and $\gamma$ a twice-symmetric, \emph{contra\-variant}, non-negative tensor field, whose kernel is generated by the nowhere vanishing  $1$-form $\theta$. 

The \emph{strong definition} requires, in addition, a 
symmetric affine con\-nection $\nabla$ with Christoffel symbols 
$\Gamma^k_{ij}$, which parallel-transports both $\gamma$ and $\theta$ \cite{Kunzle}, extending the previous triple to a quadruple 
$
(\IN,\gamma,\theta,\nabla).
$

Note that $\nabla$ is not uniquely defined by the  $(N,\gamma,\theta)$ and it is precisely this ambiguity which is responsible for the multiplicity of conformal  structures discussed in this paper.

The ``clock'' one-form is closed, $d\theta=0$, so $\ker\theta$ is thus a Fr\"obenius-integrable
 distribution, whose leaves  are $d$-dimensional and are endowed with a Riemannian structure inherited from $\gamma$ \cite{Kunzle}. The quotient 
 $\IN/\ker\theta$ 
is $1$-dimensional~: it is the absolute Newtonian time-axis (which can be either compact or non-compact). 

The standard flat NC structure is given, in an adapted coordinate system $(x^i)$, by
\begin{equation}
\IN^{d+1}=\bbR^d\times\bbR,
\qquad
\gamma=\delta^{AB}\frac{\partial}{\partial{}x^A}\otimes\frac{\partial}{\partial{}x^B},
\qquad
\theta=dt,
\qquad
\Gamma^k_{ij}=0
\label{NC}
\end{equation}
for all $i,j,k={0,1,\ldots,d}$,  where $A,B=1,\dots d$, and $t=x^{d+1}$ is the Galilean time-coordinate. (In the weak case, the
Christoffel symbols are ignored).
Other non-trivial NC structures are presented \cite{DHGalConf}. 
\goodbreak
 
The automorphisms, i.e., transformations which preserve all geometrical in\-gredients of the theory provide us with
[generalized] Galilei algebras. In the weak case, these are
typically \emph{infinite-dimensional}, but become finite dimensional when the strong definition is used \cite{GalIso}. For the flat NC structure (\ref{NC}), and without considering the flat NC connection, we get the  \emph{Coriolis Lie algebra}
\beq
X=\Big(\omega_B^A(t)x^B+\eta^A(t)\Big)\frac{\; \p}{\p x^A}+\tau\frac{\, \p}{\p t}
\label{weakGal}
\eeq
where $\omega(t)\in\so(d)$ and $\eta(t)$  are arbitrary functions of time and $\tau\in\bbR$ \cite{GalIso}, while the strong definition leaves us with the usual \textit{Galilei group} $\Gal(d+1)$,  
 represented by the  matrices, 
\begin{equation}
a =
\left(
\begin{array}{ccc}
R&\bb&\bc\\
0&1&e\\
0&0&1
\end{array}
\right)\in\Gal(d+1),
\label{Gal}
\end{equation}
where $R\in\rO(d)$, $\bb,\bc\in\bbR^d$, and $e\in\bbR$, cf. \cite{Trautman,SSD}. 
 Then the \emph{Galilei Lie algebra} $\gal(d+1)$ is isomorphic to the vector fields of $N^{d+1}$, 
\begin{equation}
X=(\omega^A_B\,x^B+\beta^A t +\gamma^A)\frac{\partial}{\partial{}x^A}+\tau\frac{\partial}{\partial{}t}\in\gal(d+1),
\label{Galmatrix}
\end{equation}
where $\omega\in\so(d)$, $\bbeta,\bgamma\in\IR^d$ and $\tau\in\IR$.

\goodbreak

\subsection{Carroll manifolds}\label{CarrollManif}

Analogous (but ``dual'') definitions of a
\emph{Carroll manifold} can be proposed  \cite{Carrollvs,DGHBMS}. The \emph{weak definition} requires having
a triple $(\IC,\rg,\xi)$ where $C$ (for Carroll) is again a smooth $(d+1)$-dimensional manifold, endowed with a twice-symmetric \emph{covariant}, positive, tensor field $\rg$, whose kernel is generated by the nowhere vanishing, complete vector field $\xi$. The \emph{strong definition} requires, in addition,  a sym\-metric affine con\-nection $\nabla$  that parallel-transports both $\rg$ and $\xi$, extending the triple to  
 a quadruple $(\IC,\rg,\xi,\nabla)$.
Note that, just as in the Galilei framework, the degeneracy of the ``metric'' $\rg$ implies that the con\-nection~$\nabla$ is not uniquely defined by the pair~$(\rg,\xi)$. 

\goodbreak

The standard weak  Carroll structure is given, in an adapted coordinate system $(x^i)$, by 
\begin{equation}
C^{d+1}=\bbR^d\times\bbR, 
\qquad
\rg=\delta_{AB}\,dx^A\otimes{}dx^B,
\qquad
\xi=\frac{\partial}{\partial s},
\label{flatC}
\end{equation}
completed, in the strong case, with
\beq
\Gamma^k_{ij}=0
\label{flatconn}
\eeq
for all $i,j,k=0,1,\ldots,d$, where $s=x^{d+1}$ is now the ``Carrollian time'' coordinate. The coordinate $s$ has the dimension of [action]/[mass].

\goodbreak

Further examples of a Carroll manifolds were discussed in Ref. \cite{DGHBMS};
see also Section \ref{Examples} below.

The isometry group of the weak
 Carrollian structure $(C,\rg,\xi)$ is infinite-dimensional, since it is invariant under the mappings
\beq
x'^A=x^A,\qquad
s' = s + f(x^1,\ldots,x^d)
\eeq
for an arbitrary function $f$.
However, requiring the preservation of the connection~$\nabla$ implies that
the automorphisms of the flat Carroll structure (\ref{flatC})  form the finite-dimensional \textit{Carroll group} \cite{Leblond, SenGupta,DGH}, which we will denote by $\Carr(d+1)$.
The latter is represented by the matrices,
\begin{equation}
a =
\left(
\begin{array}{ccc}
R&0&\bc\\
-\barbb{R}\,&1&f\\
0&0&1
\end{array}
\right)\in \Carr(d+1),
\label{Carrmatrix}
\end{equation}
where $R\in\rO(d)$, $\bb,\bc\in\IR^d$, and $f\in\IR$ represent rotations, Carrollian boosts, space translations and Carrollian time translations, respectively.
The superscript $\{\,\cdot\,\}^T$ denotes transposition.

Note the dual aspect of the matrix representation (\ref{Carrmatrix}) when compared to the Galilean case, (\ref{Galmatrix}).

The Carroll Lie algebra, $\carr(d+1)$, is isomorphic to the Lie algebra of the following vector fields of $\IC$, i.e. ,
\begin{equation}
X=(\omega^A_B\,x^B+\gamma^A)\frac{\partial}{\partial{}x^A}+(\sigma-\beta_A\,x^A)\,\frac{\partial}{\partial{}s}\in\carr(d+1),
\label{carralg}
\end{equation}
where $\omega\in\so(d)$, $\bbeta,\bgamma\in\bbR^d$, and $\sigma\in\bbR$.

\subsection{Unification: Bargmann, Newton-Cartan, Carroll}\label{Unifchap}

 
Recall that a \emph{Bargmann manifold} is a triple 
$(\IB,\BG,\xi)$,
 where $\IB$ (for Bargmann) is a $(d+2)$-dimensional 
manifold with $\BG$ a  
metric of Lorentz signature $(d+1,1)$, and the ``vertical'' vector, $\xi$, a nowhere vanishing, complete, null vector, which is parallel-transported by the Levi-Civita
con\-nection $\nabla$ of $\BG$ \cite{DBKP,DGH}.

The flat Bargmann structure is given, in an adapted coordinate system $(x^i)=(x^A,t,s)$, by 
\begin{equation}
\IB=\bbR^d\times\bbR\times\bbR, 
\quad
\BG=\sum_{A,B=1}^d\delta_{AB}\,dx^A\otimes{}dx^B + dt\otimes{}ds+ds\otimes{}dt,
\quad
\xi=\frac{\partial}{\partial s}\,.\;
\label{Bstruct}
\end{equation} 
Note that both
$s$ and $t$ are  light-cone, i.e. null, coordinates.
The automorphisms of $(B,G,\xi)$, i.e., the $\xi$-preserving isometries $a$ of the flat Bargmann structure (\ref{Bstruct}), 
\beq
a^*\BG=\BG,
\qquad
a_*\xi=\xi
\label{BargBarg}
\eeq 
form the \emph{Bargmann group} (also called extended Galilei group) $\Barg(d+1)$ \cite{DBKP,DGH}, i.e., the group of those matrices of the form
\begin{equation}
a =
\left(
\begin{array}{cccc}
R&\bb&0&\bc\\
0&1&0&e\\
-\barbb{}R\;&-\half\bb{}^2&1&f\\
0&0&0&1
\end{array}
\right)\in\Barg(d+1,1),
\label{Barg}
\end{equation}
where $R\in\rO(d)$, $\bb,\bc\in\IR^d$, and $e,f\in\IR$. 

\goodbreak

The Bargmann Lie algebra $\barg(d+1)$ is hence isomorphic to the Lie algebra of the  vector fields of $\IB$,
\begin{equation}
X=(\omega^A_B\,x^B+\beta^A\,t+\gamma^A)\frac{\partial}{\partial{}x^A}+\tau\frac{\partial}{\partial{}t}+(\sigma-\beta_A\,x^A)\,\frac{\partial}{\partial{}s}\in\barg(d+1),
\label{carr}
\end{equation}
where $\omega\in\so(d)$, $\bbeta,\bgamma\in\bbR^d$, and $\tau,\sigma\in\bbR$.

In the flat case, the Bargmann group (\ref{Barg}) is a non-trivial central extension of the Galilei group
 with ``vertical'' translations generated by $\xi=\p/\p{}s$.
 
\goodbreak
 Factoring out flat Bargmann space, $B$, by the ``vertical'' translations generated by $\xi$, the quotient, $N=B/\bbR$, acquires a (flat) NC structure \cite{DBKP,DGH}.
 Let us call  $\vartheta=\BG(\xi)$ the $1$-form associated to $\xi$ on the general Bargmann manifold $(\IB,\BG,\xi)$. 
Since $L_\xi\,\BG=0$, the contravariant symmetric $2$-tensor $\BG^{-1}$ projects to $\IN$  as the (rank $d$) contravariant tensor field 
$\gamma$. 
 Similarly,  $\vartheta=\BG(\xi)$ is the pull-back to $\IB$ of a ``clock'' $1$-form $\theta$ on the quotient.

Finally, it has been shown that the Levi-Civita connection, $\nabla$, of $\IB$ naturally defines an affine connection $\nabla^{\IN}$ on $\IN$ that transports $(\gamma,\theta)$. A Bargmann structure thus projects onto the strong NC structure $(\IN,\gamma,\theta,\nabla^{\IN})$. See~\cite{DBKP}.


Now consider on $\IB$ the $(d+1)$-dimensional distribution defined by
$
\ker\vartheta
$, which is indeed
 the orthogonal complement of $\xi$, and  is, again, integrable since $d\vartheta=0$. 
  The ``clock'' $1$-form, $\theta$ is locally 
  $\vartheta=dt$. 
Notice that the  ``vertical'' vector field $\xi$  belongs to this foliation, since $\vartheta(\xi)=\BG(\xi,\xi)=0$.
Call 
\begin{equation}
\iota:\IC\hookrightarrow{}\IB
\label{Ct}
\end{equation}
the embedding  of a leaf of $\ker\vartheta$, at at $t=0$, say.
Then a routine computation shows that the embedding
(\ref{Ct}) endows $\IC$ with a $(d+1)$-dimensional Carroll structure. 
 The  flat Bargmann structure (\ref{Bstruct})  
yields, in particular, the standard flat Carroll structure (\ref{flatC}).

Endowing $\IC$ with the induced symmetric covariant $2$-tensor $ \rg^{\IC}=\iota^*\BG,$ the quadruple
 $(\IC,\rg^{\IC},\xi;\nabla^C)$, 
 where $\nabla^C$ is the induced connection, becomes a Carroll manifold in the sense of Section \ref{CarrollManif}. 
The flat Bargmann structure (\ref{Bstruct}) 
yields the standard flat Carroll structure (\ref{flatC}) \cite{Carrollvs}.
In what follows  the superscript $\{\,\cdot\,\}^{\IC}$ will be dropped.

\section{Conformal extensions}\label{ConfExt}

\subsection{Conformal Galilei groups}\label{confGal}

Recall that in the relativistic case, where the metric $\BG$ of space-time (e.g., of Minkowski space-time $\rE^{d,1}$~), is invertible, an
infinitesimal transformation which preserves the conformal class $[\BG]$ of the metric $\BG$ is a vector field $X$ such that 
\beq
L_X[\BG]=0
\quad\Longleftrightarrow\quad
L_X(\BG\otimes \BG^{-1})=0.
\label{relconf}
\eeq
Below, we will adapt this defining property of infinitesimal conformal trans\-formations to both the Galilean and Carrollian cases.


Let us first remind the reader of the Galilean case \cite{DHGalConf,Carrollvs}. There we start with a [weak] Newton-Cartan space-time $(\IN,\gamma,\theta)$
with degenerate ``metric'' $\gamma$ and ``clock''~$\theta$. We choose a positive integer $\cN$ and call
 a (local) diffeomorphism, $a$, of $(N,\gamma,\theta)$ 
 a \emph{conformal Galilei transformation of level 
$\cN\in\bbN$} if
\begin{equation}
a^*\big(\gamma\otimes\theta^\cN\big)=\gamma\otimes\theta^\cN,
\label{ConfGalN}
\end{equation}
where $\theta^k$ is a shorthand for the $k$-th tensor power $\theta^{\otimes{}k}$. 

Infini\-tesimally, the \textit{conformal Galilei algebra of level $\cN$}, we denote by $\cgal_\cN(N,\gamma,\theta)$ is therefore spanned by the vector fields $X$ which are solutions of
\begin{equation}
L_X\big(\gamma\otimes\theta^\cN\big)=0.
\label{CGAN}
\end{equation}
Spelt out in separate terms, (\ref{CGAN})
amounts to requiring
\begin{equation}
L_X\gamma=\lambda\,\gamma
\qquad
\&
\qquad
L_X\theta=\mu\,\theta
\qquad\hbox{such that}\qquad
\lambda+\mu\,{\cN}=0.
\label{CGANbis}
\end{equation}

In the flat case, the corresponding Lie algebra is spanned by the vector fields
\begin{equation}
X=
\Big(
\omega^A_B(t)x^B+\frac{\cN}{2}T'(t)x^A+\eta^A(t)
\Big)\frac{\partial}{\partial x^A}
+
T(t)\frac{\partial}{\partial t}\in\cgal_\cN(d+1)\,,
\label{cgalz} 
\end{equation}
where $\bomega(t)\in\so(d)$, $\veta(t)$ and $T(t)$, interpreted as (generalized) rotations, translations and time-reparametrizations, depend arbitrarily on time, $t$. 
This Lie algebra is  infinite-dimensional, and generates  $\CGal_{\cN}$, the Conformal Galilei group of level $\cN$.
 
Demanding the strong definition, i.e., that these transformations should also preserve the projective structure associated with some  NC connection $\nabla$, we end up with a finite-dimensional group $\pSch_\cN(N,\gamma,\theta,\nabla)$ or, if the transformation acts improperly, with a much larger pseudo-group. See  \cite{DHGalConf,DLazzari} and references therein.

\goodbreak

In particular, the choice $\cN=1$ yields the infinite-dimensional  \textit{Schr\"odinger-Virasoro algebra}, \cite{Henkel94,RU} denoted by $\sv(d+1)$ in the flat NC case; see also \cite{DHGalConf}. Its intersection with the [infinitesimal] projective group [i.e., (local) diffeomorphisms of $(N,\gamma,\theta)$ that permute the un\-para\-metrized geodesics of $\nabla$] is the (finite dimensional) \textit{Schr\"odinger Lie algebra} of $(\IN,\gamma,\theta,\nabla)$. In the flat NC case, this algebra, 
$$
\sch(d+1)=\cgal_1(d+1)\cap\Sl(d+2),
$$
features a dynamical exponent $z=2$. 

\goodbreak

For $\cN=2$, which, as said above,  extends the Wigner-In\"on\"u contraction from Poincar\'e  to Galilei we get 
 what is called simply ``the'' \textit{Conformal Galilei Lie Algebra}, 
$$
\cga(d+1)=\cgal_2(d+1)\cap\Sl(d+2);
$$
it has dynamical exponent $z=1$
\cite{GalConf,DHGalConf}.
 
Let us spell out, for further use, the $(1+1)$-dimensional flat case,
\beq
N=\bbR\times{}S^1
\qquad
\gamma=\frac{\p}{\p{}x}\otimes\frac{\p}{\p{}x},
\qquad
\theta =dt,
\qquad
\Gamma_{ij}^k=0,
\label{1+1NC}
\eeq
where $t$ is an affine time-coordinate on $S^1$.
(If we want to produce the Virasoro cocycle, the ``time axis'' must be compact.) The
 Conformal Galilei Algebra (CGA) is  spanned by the vector fields
\begin{equation}
X=
\big(\alpha'(t)x+\eta(t)\big)\frac{\partial}{\partial x}
+\alpha(t)\frac{\partial}{\partial t}\in\cgal_2(1+1).
\label{cga1+1} 
\end{equation}

This Lie algebra admits the centre-free Virasoro algebra as the Lie subalgebra genera\-ted by the 
vector field $\alpha(t)\,\p/\p{t}$
on $S^1$.
If $X_3=[X_1,X_2]$, we find 
$$\alpha_3=\alpha_1\alpha'_2-\alpha_2\alpha'_1
\quad
\hbox{and}
\quad 
\eta_3=(\alpha_1\eta'_2-\eta_2\alpha'_2)+(\eta_1\alpha'_2-\alpha_2\eta'_1);
$$
this corresponds to Eq.~(2.4) of Bagchi et al. in Ref. \cite{Bag1}. This implies that 
\beq
\cgal_2(1+1)\cong\Vect(S^1)\ltimes{}C^\infty(\bbR).
\eeq 

In \cite{Bag1} it has been shown how $\cgal_2(1+1)$ arises as a ``non-relativistic contraction'' of two (centre-free) Virasoro algebras, i.e.  of holomorphic vector fields of the complex plane.
The Lie algebra $\cgal_2(1+1)$ clearly admits a non-trivial central extension by~$\bbR$, the Virasoro cocycle of the Lie subalgebra $\Vect(S^1)$ being constructed via the infinitesimal Schwarzian derivative \cite{OT}
$
s(\alpha(t)\partial/\partial{}t)={\alpha}'''(t)\,dt^2.
$
The reader is referred to \cite{GalConf,DHGalConf} for the general finite-dimensional case with prescribed dynamical exponent $z=2/\cN$. Central extensions of the Virasoro algebra were classified in \cite{Ovsienko}.

It is worth mentioning that requiring in addition 
that the clock  $\theta$ be  preserved,
the conformal Galilei group with  $\cN=0$ is  the Coriolis group (\ref{weakGal}) of Galilean isometries.
The affine Coriolis transformations, i.e., such that $a^*\nabla=\nabla$, then form the 
 \textit{Galilei group} of $(\IN,\gamma,\theta,\nabla)$ \cite{Trautman}.

\bigskip

\subsection
{The (extended) Schr\"odinger group}\label{SchSection}
 
The ``$\cN=1$ conformal Galilei group'' is just the 
(extended) \emph{Schr\"odinger group}, which can also be treated in 
the Bargmann framework outlined in Section \ref{Unifchap}.
In fact it is  the group of all conformal transformations $a$ of~$(\IB,\BG,\xi)$ that also preserve the ``vertical'' vector field $\xi$, 
\beq
a^*\BG=\Omega^2\BG
\qquad
\&
\qquad
a_*\xi=\xi
\label{SchBarg}
\eeq 
for some strictly positive function $\Omega$ of $\IB$ depending on $a$.  
The (pseudo-)group, $\widetilde{\Sch}(B,G,\xi)$, of those transformations satisfying (\ref{SchBarg}) projects down to NC space-time as the Schr\"odinger group $\Sch(N,\gamma,\theta,\nabla)$, and turns out to be a one-dimensional central extension of the latter \cite{DGH}. 
 As a group of conformal transformations, the (extended) Schr\"odinger group is finite-dimensional. 
 
\goodbreak
In the flat case, this group descends to NC space-time as the group of matrices, 
\begin{equation}
a =
\left(
\begin{array}{ccc}
R&\bb&\bc\\
0&d&e\\
0&f&g
\end{array}
\right)\in\Sch(d+1),
\label{Sch}
\end{equation}
where $R\in\rO(d)$, $\bb,\bc\in\bbR^d$, and $d,e,f,g\in\IR$ with $dg-ef=1$.
We record for further reference that its Lie algebra $\sch(d+1,1)$ is spanned by the vector fields $X$ solutions of (\ref{sch}), namely
\begin{eqnarray}
X
&=&
\nonumber
\left(\omega^A_B\,x^B+\beta^At+\gamma^A+2\alpha{}tx^A+\chi{}x^A\right)
\frac{\partial}{\partial{}x^A}
+
\left(2\kappa{}t^2+2\chi{}t+\tau\right)
\frac{\partial}{\partial{}t}
\\
&&+\left(-\bbeta\cdot\bx-\kappa\,\bx^2+\varphi\right)
\frac{\partial}{\partial{}s}\in\sch(d+1,1)
\label{schbis}
\end{eqnarray}
where $\bomega\in\so(d)$, $\bbeta,\bgamma\in\bbR^d$, $\kappa,\chi,\tau,\varphi\in\bbR$.

Equation (\ref{SchBarg}) is
clearly  a conformal extension of the Bargmann group   
(\ref{BargBarg}). See also (\ref{Sch}) and (\ref{Barg}) for the flat case.

Let us mention that this treatment of the Schr\"odinger case is unique in that the other Conformal Galilei groups, i.e., those with $\cN>1$, admit 
\emph{no lift} to Bargmann space.

\goodbreak


\subsection{Conformal Carroll transformations
}\label{cNconfCarr}

Now we  define 
 \emph{conformal Carroll transformations of level  $\cN$}, we denote by $\confcarr_{\cN}(d+1)$. Our definition relies on  proper
Carroll structures and the ``dual'' nature of the 
latter to those of Newton-Cartan. 
Trading the ``clock'' $1$-form
$\theta$ in (\ref{ConfGalN}) for its ``dual'' object,
i.e., the ``vertical'' vector $\xi$, we consider
the trans\-formations $a$ of $C$ which satisfy 
\begin{equation}
a^*\big(\rg\otimes\xi^{\cN}\big)
=
\rg\otimes\xi^{\cN}.
\label{CCN}
\end{equation}
 The (pseudo-)group of conformal Carroll transformation of level  $\cN$ 
  will be denoted by $\ConfCarr_{\cN}(C,\rg,\xi)$.
  
\goodbreak
  
 Infinitesimally, we translate (\ref{CCN}) as
\beq
L_X\big(\rg\otimes\xi^{\cN}\big)=0,
\label{cCN}
\end{equation}
which amounts to requiring
\begin{equation}
L_X\rg=\lambda\,\rg
\qquad
\&
\qquad
L_X\xi=\mu\,\xi
\qquad
\hbox{with}
\qquad
\lambda+\cN\mu=0,
\label{ccarrNinit}
\end{equation}
cf. (\ref{CGANbis}).
The Lie algebra of infinitesimal conformal Carroll transformation of level~$\cN$, i.e., of all solutions $X$ of the system (\ref{cCN}) will be denoted by $\confcarr_\cN(C,\rg,\xi)$.


For the flat Carroll structure (\ref{flatC}), i.e.,
$\IC=\bbR^{d+1}$, with 
$\rg=\delta_{AB}\,dx^A\otimes{}dx^B$,  
$\xi=\partial/\partial{s}$,
the vector fields $X=X^A\,\partial/\partial x^A+T\,\partial/\partial s$ satisfying (\ref{ccarrNinit}) are such that 
$$
\partial_AX_B+\partial_BX_A=\lambda\,\delta_{AB},
\quad 
\partial_sX^A=0,
\quad
\partial_sT=-\mu
$$
for all $A,B=1,\ldots,d$.   
For $\cN\neq0$, our vector fields are of the form
\begin{eqnarray}
\nonumber
X&=&\left(\omega^A_B\,x^B+\gamma^A+\chi{}x^A+\kappa^Ax_Bx^B-2\kappa_Bx^Bx^A\right)\frac{\partial}{\partial{}x^A}
\\[6pt]
&&+\left(\frac{2}{\cN}(\chi-2\kappa_Ax^A)\,s+F(x^A)\right)\frac{\partial}{\partial{}s}\in\confcarr_{\cN}(d+1)
\label{confcarrN}
\end{eqnarray}
with $\bomega\in\so(d)$, $\bgamma,\bkappa\in\bbR^d$, $\chi\in\bbR$, and $F\in{}C^\infty(\bbR^d,\bbR)$;
they generate the Lie algebra we have already denoted by 
$\confcarr_{\cN}(d+1)$ in (\ref{confcarrN}). The conformal factor in (\ref{ccarrNinit}) is
expressed as
\beq
\lambda=2(\chi-2\kappa_Ax^A).
\eeq

Our Lie algebra is infinite-dimensional, since $F$ is an arbitrary function of the ``space'' variables $x^A$.
The Carroll Lie algebra $\carr(d+1)$ itself corresponds to the Lie subalgebra obtained by
 choosing
$ 
 \chi=\kappa=0,\;
 F=\sigma-\beta_Ax^A,
$ 
as in (\ref{carralg}).

The two following extreme cases are particularly interesting.

\goodbreak

\bigskip


(i) For $\cN=0$, the defining condition 
(\ref{CCN}) becomes
\begin{equation}
L_X\rg=0
\qquad
\&
\qquad
L_X\xi=\mu\,\xi\,,
\label{ccarrzero}
\end{equation}
and we end up with the algebra we call $\confcarr_0(d+1)$, which is spanned by the vector fields
\begin{eqnarray}
X&=&\left(\omega^A_B\,x^B+\gamma^A\right)\frac{\partial}{\partial{}x^A}
+T(x^A,s)\,\frac{\partial}{\partial{}s}
\in\confcarr_0(d+1),
\label{confcarr0}
\end{eqnarray}
where $(\bomega,\bgamma)\in\mathfrak{e}(d)$, and $T\in{}C^\infty(\bbR^{d+1},\bbR)$. 

\goodbreak

It follows that 
$\confcarr_0(d+1)$ 
 is the semi-direct product of the  Euclidean Lie algebra with ``supertranslations'' of Carrollian ``time'',
\begin{equation}
\confcarr_0(d+1)\cong\mathfrak{e}(d)\ltimes{}C^\infty(\bbR^{d+1},\bbR).
\label{isominfty}
\end{equation}
The terminology will be explained in Example 3. of Sec \ref{Examples}.


(ii) Taking instead the limit $\cN\to\infty$, from 
 (\ref{ccarrNinit}) we infer,
 using (\ref{confcarrN}),  the  explicit expression valid in the flat case, 
\begin{eqnarray}
\nonumber
X&=&\left(\omega^A_B\,x^B+\gamma^A+\chi{}x^A+\kappa^Ax_Bx^B-2\kappa_Bx^Bx^A\right)\frac{\partial}{\partial{}x^A}\\
&&+T(x^A)\,\frac{\partial}{\partial{}s}\in\confcarr_\infty(d+1).
\label{confcarrinfty}
\end{eqnarray}
Hence
\begin{equation}
\confcarr_\infty(d+1)\cong\so(d+1,1)\ltimes{}C^\infty(\bbR^d,\bbR)
\label{isominfty1}
\end{equation}
which is infinite-dimensional because of the presence of ``supertranslations'', given by the function $T(x^A)$.

Let us observe that the transformations $a$ of $\IC$ generated by 
(\ref{confcarrinfty}) satisfy \emph{formal\-ly} the same conditions (\ref{SchBarg}) as those defining the Schr\"odinger group, but for mappings of  the Carroll manifold $(\IC,\rg,\xi)$, and with
the Bargmann metric $\BG$  replaced by the degenerate Carroll ``metric'' $(\rg,\xi)$, namely
\begin{equation}
a^*\rg=\Omega^2\rg
\qquad
\&
\qquad
a_*\xi=\xi
\label{ConfCarrDef}
\end{equation}
for some strictly positive function $\Omega$ of $C$, depending on $a$.
The infinitesimal version of (\ref{ConfCarrDef}) is
\begin{equation}
L_X\rg=\lambda\,\rg
\qquad
\&
\qquad
L_X\xi=0
\label{standconfcarr}
\end{equation}
for some function $\lambda$ on $\IC$.  The solutions of (\ref{standconfcarr}) reproduce in fact (\ref{confcarrinfty}).

\subsection{Conformal Carroll subgroups of the Schr\"odin\-ger group}\label{SchCarr}

 The (pseudo-)group of conformal transformations of
Bargmann space is finite-dimensional and the subgroup which preserves the ``vertical'' vector $\xi$ form  the Schr\"odinger group \cite{DBKP,DGH,DLazzari}. But the Carroll manifold 
can be embedded into Bargmann space as a $t=\const$ slice and one may wonder what  part of the Schr\"odinger symmetry is consistent  with the  constraint $t=\const$ Then a calculation (not detailed here) shows that, in the flat case, we are left with the
Schr\"odinger-Carroll algebra $\scarr(d+1)$
\begin{eqnarray}
X=(\omega^A_B\,x^B+\gamma^A+\chi{}\,x^A)\frac{\partial}{\partial{}x^A}
+\big(\sigma-\beta_A\,x^A-\half\kappa\,x_Ax^A\big)\,\frac{\partial}{\partial{}s},
\label{ccarr}
\end{eqnarray}
where $\bomega\in\so(d)$, $\bbeta,\bgamma\in\bbR^d$ and $\chi,\kappa,\sigma\in\bbR$, which is clearly an extension of the Carroll action (\ref{carralg}) with space but no ``time'' (i.e., $s$) dilations $\chi$. The supertranslations are time-independent and
 also involve Schr\"odinger expansions with $\kappa$, namely
\beq
T(x^A)=\sigma-\beta_A\,x^A-\half\kappa\,x_Ax^A.
\eeq 
A straightforward compu\-tation shows that the Lie algebra (\ref{ccarr}) is 
\begin{equation}
\scarr(d+1)=
\Big\{X\in\confcarr_\infty(d+1)\;\strut\big\vert\;\partial_a\partial_bX^c+\alpha\,\rg_{ab}\,\xi^c\equiv0,\; \alpha\in\bbR \Big\}\;.
\label{ccarrBis}
\end{equation}
For a general Carroll manifold $(C,\rg,\xi)$ embedded into Bargmann space, our algebra 
 satisfies
\begin{equation}
\scarr(C,\rg,\xi,\nabla)=
\left\{X\in\confcarr_\infty(C,\rg,\xi)\;\strut\big\vert\; L_X\nabla+\alpha\,\rg\otimes\xi=0,\; \alpha\in\bbR\right\},
\label{ccarrTer}
\end{equation}
where  $\nabla$ is the induced Levi-Civita connection from Bargmann space.

\goodbreak

\section{Examples of Carroll spacetimes \&   symmetries}\label{Examples}

We now illustrate our general theory by some selected examples.

\begin{enumerate}
\item
Let us first consider  $(1+1)$-dimensional flat Carroll space-time,
\beq
C=\IR\times\IR,
\quad
\rg=dx^2,
\quad
\xi=\frac{\p}{\p{}s},
\quad\Gamma_{ij}^k=0,
\label{1+1flatCarroll}
\eeq
to find the $\cN$-conformal Carroll algebra spanned by the vector fields
\begin{equation}
X=
\alpha(x)\frac{\partial}{\partial x}
+
\Big(\frac{\cN}{2}\alpha'(x)s+\eta(x)\Big)
\frac{\partial}{\partial s}\in\confcarr_\cN(1+1),
\label{cCn1+1} 
\end{equation}
where $\alpha$ and $\eta$ are arbitrary smooth functions of ``space'' $\bbR$.

For the ``relativistic'' value $\cN=2$, this reduces precisely to the Conformal Galilei case (\ref{cga1+1}), whenever  \emph{Newtonian time is changed into position, and position into Carrollian time},
\beq
t \to x
\qquad
\&
\qquad
x\to s,
\label{txs}
\eeq
confirming that the two algebras do indeed have the same structure \cite{Bag1,Bag2}.
Note  that
(\ref{txs}) also swaps Galilean and Carrollian boosts.

\item
CFT-type applications envisaged in \cite{Bag1,Bag2} 
require to work in the plane, $d=2$, e.g., with
\beq
C
=\bbR^2\times\bbR,
\quad
\rg=dx^2+dy^2,
\quad
\xi=\frac{\p}{\p{}s},
\quad
\Gamma_{ij}^k=0.\;
\eeq

\goodbreak

In  geometric terms, the conformal $\cN=\infty$ Carroll Lie algebra is spanned by the vector fields 
\begin{equation}
X=h(z)\,\p_z+\overline{h(z)}\,\p_{\bar{z}}+f(z)\,\p_s\in\confcarr_{\infty}(1+1),
\label{scC2+1}
\end{equation}
where $h(z)$ is a holomorphic, and $f(z)$ a smooth real valued function, on the complex plane.
Our $\cN$-conformal Carroll vectors fields are, in turn,
\begin{equation}
X=h(z)\,\p_z+\overline{h(z)}\,\p_{\bar{z}}+\left(\frac{s}{\cN}
\big(\p_zh+\p_{\bar{z}}\bar{h}\big)
+f(z)\right)\p_s \in\confcarr_\cN(1+1).
\label{cCN2+1}
\end{equation}
Letting $\cN\to\infty$ the
 conformal Carroll algebra $\ccarr_\infty$ is recovered.
Again, we have a $1$-dimensional central extension of $\confcarr(1+1)$ governed by the infinitesimal Schwarzian derivative
$
s(h(z)\partial/\partial{}z)=h'''(z)\,dz^2.
$
 
\item
More general Carroll manifolds are given by
\begin{equation}
C=\Sigma\times\bbR,\;
\rg=\hrg_{AB}(x)\,dx^A\otimes{}dx^B,\;
\xi=\frac{\partial}{\partial s},\;
\Gamma_{AB}^C=\hGamma_{AB}^C,\;
\Gamma_{AB}^s=\Gamma_{AB}
\label{Cbis}
\end{equation}
where $(\Sigma,\hrg)$ is a $d$-dimensional Riemannian manifold, with Christoffel symbols~$\hGamma_{AB}^C$, and where the symmetric quantities $\Gamma_{AB}$ remain arbitrary.

Computing the general expression for a conformal Carroll vector field of level~$\cN$ yields 
\beqa
 X=\hX+
\Big(\frac{2\lambda}{\cN{}\,d}\,s
+
F(x^A)\Big)\frac{\partial}{\partial{}s}\,,
\label{ccSigma}
\eeqa
where $\hX=\hX^A(x)\,\partial/\partial{x^A}$
is a conformal vector field of $(\Sigma,\hrg)$, i.e., such that 
\beq
L_{\hX}\,\hrg=\lambda\,\hrg
\quad\hbox{with} \quad
\lambda=\frac{2}{d}\,\widehat\nabla_A\hX^A,
\label{lambdaSigma} 
\eeq
and $F$ is a density  on $\Sigma$ of weight $-2/(\cN{}\,d)$, i.e,  is a real function  $F$ which transforms as $F\to \Omega^{-2\cN}F$ under a  rescaling $\hrg\to\Omega^2\hrg$ of the metric.

Integration of the vector field (\ref{ccSigma})  yields the group action $(x,s)\mapsto(x',s')$, 
where 
\beq
x'=\varphi(x), 
\quad
s'=\Omega^{2/N}(x)\big(s+\alpha(x)\big),
\label{TSigma}\eeq
with $\varphi\in\Conf(\Sigma,\hrg)$, and $\alpha\in{}C^\infty(\Sigma,\bbR)$.

\bigskip
For example, $\Sigma$ could be the \emph{circle},
$\Sigma=\IS^1$.
Then we would get, for level-$\cN$ conformal Carroll transformations, the semi-direct product of the conformal transformations of $S^1$, namely $\Diff(S^1)$, with supertranslations of weight $-2/\cN$. They are generated by the vector fields 
\beq
X=\hX(\theta)\frac{\partial}{\partial\theta}+\Big(\frac{2}{\cN}\hX'(\theta)\,s+F(\theta)\Big)\frac{\partial}{\partial{s}}.
\eeq

Similarly, consider 
the two-sphere, $\Sigma=\IS^2$,  endowed with its standard round metric.
The conformal Carroll vector fields of level $\cN$ is given by (\ref{ccSigma}) with $d~=~2$. Recalling that the conformal transformations of the two-sphere constitute the identity component of the Lorentz group, 
$\PSL(2,\bbC)=
\SL(2,\bbC)/\bbZ_2$, we conclude that the conformal Carroll transformations are now the semi-direct product of $\SL(2,\bbC)$ with supertranslations, and hence
$$
\confcarr_{\cN}(S^2\times\bbR,\rg,\xi)
\cong
\mathfrak{sl}(2,\bbC)\ltimes{}C^\infty(S^2,\bbR).
$$
In particular, for $\cN=2$ we get, using (\ref{lambdaSigma}),
\beqa
 X=\hX+
\Big(\frac12\widehat\nabla_A\hX^A\,s+F(x^A)\Big)\frac{\partial}{\partial{}s}\,,
\label{ccS2}
\eeqa
which is, indeed the BMS Lie algebra \cite{BMS,Barnich}.
 We recall that
the BMS group is an infinite dimensional extension
of the Poincar\'e group which arises as the asymptotic symmetry group of asymptotically flat four dimensional spacetimes \cite{BMS}. The Poincar\'e group is the semi-direct product of the Lorentz group with
the four-dimensional abelian group of translations;
the BMS group is the semi-direct product of the Lorentz group with an infinite dimensional abelian group,
which may be realized as functions of a certain conformal
weight on the 2-sphere. Since ordinary translations are
realized as functions spanned by the lowest four harmonics
on the 2-sphere, members of this infinite dimensional abelian group are referred to as \emph{super-translations}. 

\item
Now we choose $C$ to be the punctured future light-cone in Minkowski space $\bbR^{d+1,1}$. The ``metric'' $\rg$ is simply the induced Minkowski metric, and $\xi$ the restriction to $C$ of the Euler vector field of $\bbR^{d+1,1}$. We have $C\cong{}S^d\times\bbR^{>0}$, which is described by those $t(\bu,1)\in\bbR^{d+1,1}$ where $\bu\in{}S^d$ and $t>0$. Then $\rg=t^2\,\hrg$ where $\hrg$ is the round metric of $S^d$, and $\xi=t\,\partial_{t}$. We will also put $s=\log t$ for convenience.

Hence the light-cone is an intrinsical\-ly defined Carroll manifold in the weak sense. A remarkable fact is that it carries \emph{no}  compatible  connection.
 To see this, choose the coordinate system $(x^A,s)$ on $C$, such that
\begin{equation}
g_{AB}=e^{2s}\,\hrg_{AB},
\quad
g_{As}=0,
\quad
g_{ss}=0,
\quad
\xi^A=0,
\quad
\xi^s=1,
\label{gxi}
\end{equation}
for all $A,B=1,\ldots,d$.  
Then if $\nabla$ were such a connection, then we would have
\begin{eqnarray}
\nonumber
(\nabla_sg)_{AB}
&=&\partial_s(e^{2s}\hg_{AB})-\Gamma^C_{sA}\,g_{CB}-\Gamma^C_{sB}\,g_{AC}
\\[4pt]
\label{1}
&=&2g_{AB}-\Gamma^C_{sA}\,g_{CB}-\Gamma^C_{sB}\,g_{AC}=0,
\end{eqnarray}
as well as
\begin{eqnarray}
\nonumber
(\nabla_Ag)_{sB}
&=&\partial_A\,g_{sB}-\Gamma^C_{As}\,g_{CB}-\Gamma^\bullet_{AB}\,g_{s\bullet}
\\[4pt]
\label{2}
&=&0-\Gamma^C_{As}\,g_{CB}-0=0.
\end{eqnarray}
Eq. (\ref{2}) yields trivially $\Gamma^C_{As}\,g_{CB}=0$ for all $A,B=1,\ldots,d$. Then from Eq.~(\ref{1}) we infer, using the symmetry of the connection, that $g_{AB}=0$, a contradiction. 

Luckily enough, our weak definition of a conformal Carroll transformation does not involve any connection,  and we find that $L_X\rg=\lambda\,\rg$ requires that $X=X^A\,\p/\p{}x^A+T\,\p/\p{}s$ be such that $\partial_sX^A=0$ as well as
$L_{\hX}\,\hrg=(\lambda-2X^s)\,\hrg$ where $\hX=X^A(x)\,\partial/\partial{x^A}$. 
Using that Eq. (\ref{ccarrNinit}) implies
$L_X\xi=-(\lambda/\cN)\xi$  allows us to deduce that the conformal Carroll algebra 
of the punctured future light cone is for $\cN=2$, the 
\emph{BMS algebra} (\ref{ccS2}).

As to Carroll isometries, the condition  
$\lambda=0$ would
\emph{fix the supertranslations} as 
\beq
T=-\frac{L_{\hX}\,\hrg}{2\,\hrg}
\eeq
 while leaving us with the vector field $\hX$ as a conformal vector field of $(\Sigma,\hrg)$; the Carroll ``isometries'' of the light-cone therefore span the conformal group, $\rO(d+1,1),$ of the celestial sphere, $S^d$, with rigidly fixed ``compensating'' super\-translations.

\end{enumerate}

\section{Newman-Unti group}\label{NUsec}

The Newman-Unti (NU)  group \cite{PenroseRindler} is spanned by those (local) diffeo\-morphisms~$a$ of~$C$ which preserve the  degenerate ``metric'', $\rg$, \emph{up to a conformal factor}, namely
\beq
a^*[\rg]=[\rg]. 
\eeq
This implies that the direction of $\xi$ is automatically preserved (since $a_*\xi$ lies again in the kernel of $\rg$). Its Lie algebra consists, hence, of all vector fields~ $X$ on~$C$ such that
\beq
L_X\rg=\lambda\,\rg,
\eeq
 the condition $L_X\xi=\mu\,\xi$ being automatically satisfied.
 
In the product case $C=\Sigma\times\bbR$ of Example 3. in Section \ref{Examples}, we find that 
\begin{equation} 
X=\hX+T(x^A,s)\frac{\partial}{\partial{}s},
\label{nu}
\end{equation}
with
$\hX=\hX^A(x)\,\partial/\partial{x^A}$ a conformal vector field of $\Sigma$, and 
$ 
T\in{}C^\infty(C,\bbR)
$ 
now an arbitrary function of $x^A$ and $s$. 
Therefore the Newman-Unti group is, 
\begin{equation}
\NU(C,G,\xi)\equiv\Conf(\Sigma)\ltimes{}C^\infty(C,\bbR).
\label{NUgr}
\end{equation} 

Choose an integer $n=0,1,\dots$. An ``intermediate''  Lie subalgebra $\mathfrak{nu}_n(d~+~1)$ of the NU Lie algebra, $\mathfrak{nu}(d+1)$, defined by
\begin{equation}
L_X\rg=\lambda\,\rg,
\quad\hbox{and}\quad 
(L_\xi)^nX=0
\label{interNU} 
\end{equation}
consists of vector fields 
$X=\hX^A(x)\,\p/\p{x^A}+T\,\partial/\partial{}s$
such that
\beq
\hX\in\conf(\Sigma)
\qquad
\hbox{and}
\qquad
(\partial_s)^nT=0,
\eeq 
i.e., such that the supertranslation is a polynomial of degree $n-1$ in $s$,
\beq
T(x^A,s)=\sum_{m=1}^ns^{m-1}T_m(x^A),
\eeq
where the $T_m$ remain arbitrary functions on $\Sigma$.

Referring to Eqs (\ref{ccSigma}) and (\ref{nu}), which give the generators of the conformal Carroll Lie algebras, we can highlight 
the interesting array of nested Lie groups
\beq
\NU_1\subset\BMS\subset\NU_2\subset\cdots\subset\NU.
\label{nested}
\eeq 
We finally notice that $\NU_1=\CCarr_\infty$.

\section{Carrollian ``photons''}\label{masslessSection}

 In special relativity the massless representations of the Poincar\'e group may 
  describe  massless particles, e.g.,  photons. Massless representations  have already been investigated for the Galilei group, leading to classical models of both spinning and spinless light rays \cite{SSD,SpinOptics}. These models were constructed using coadjoint orbits of the \emph{Euclidean group}, which is in fact  a subgroup of the Galilei group. Below we pursue this idea
 for the Carroll group. We begin by reviewing its coadjoint orbits.
 
The general idea \cite{SSD} is to consider an ``evolution space'', i.e., a manifold $V$ equipped with a closed $2$-form $\sigma$ whose kernel $\ker\sigma$ has constant rank. Now, $\sigma$ being closed, the distribution $\ker(\sigma)$ is integrable; its space of leaves, $U$,
when a  manifold is endowed with a symplectic $2$-form $\omega$, given by the projection of $\sigma$. 
Souriau calls~$(U,\omega)$ the ``space of motions''.

Given a vector field $X$ on $V$ such that
$L_X\sigma=0$, we may define a function $\mu_X$ by 
\beq
d\mu_X=-\sigma(X,\,\cdot\,)
\label{dmuX}
\eeq
provided $\sigma(X,\,\cdot\,)$ is globally exact.
It follows that for all $Y\in\ker\sigma$  we find that $Y(\mu_X)=0$. Thus $\mu_X$ is a \emph{constant of the motion} in the sense that it is constant along the leaves of $\ker\sigma$; this is the \textit{presymplectic Noether theorem} \cite{SSD}. 

In the special case where $\sigma$ is exact, i.e., $\sigma=d\varpi$, and $L_X\varpi=0$, then we obtain
\beq
\mu_X=\varpi(X)
\label{muX}
\eeq
thus fixing the overall additive constant present in the general definition (\ref{dmuX}) of the \textit{momentum mapping} $\mu$, defined by $\mu_X=\mu\cdot{}X$.
 
A particular case of this general procedure is to take $V$ to be a Lie group $G$ whose Lie algebra is denoted $\fg$. 
For each $\mu_0\in\fg^*$ we define $\sigma$ by
\begin{equation}
\sigma=d\varpi
\quad
\hbox{with}
\quad
\varpi=\mu_0\cdot\Theta
\label{varpi}
\end{equation}
where $\Theta=``g^{-1}dg''$ is the left-invariant $\fg$-valued  Maurer-Cartan $1$-form on $G$.
The space of motions $U$ is then given by the coadjoint orbit $\cO_{\mu_0}=\Coad(G)\mu_0$ passing through
$\mu_0$. 

Applied to our case, we represent the Carroll Lie algebra $\carr(d+1)$ by the matrices
\begin{equation}
Z =
\left(
\begin{array}{ccc}
\bomega&0&\bgamma\\
-{\bbeta}^T&0&\alpha\\
0&0&0
\end{array}
\right)
\label{Z}
\end{equation}
with $\bomega\in\so(d)$, $\bbeta,\bgamma\in\bbR^d$, and $\alpha\in\bbR$, cf. Eq. (\ref{Carrmatrix}). Bearing in mind that Carrollian time, $s$, has  dimension  $L^2/T$, i.e., action/mass, we find that $[\bomega]=1$, $[\bbeta]=$ L/T, $[\bgamma]=$ L , and $[\alpha]=[s]$. Then an element in the dual of the Lie algebra  is
$
\mu=(\belle,\bk,\bp,m)\in\carr(d+1)^*,
$
where the pairing between the Lie algebra and its dual is defined as
\begin{equation}
\mu\cdot{}Z=\half\Tr(\belle\bomega)-\bk\cdot\bbeta-\bp\cdot\bgamma+m\alpha.
\label{muZ}
\end{equation}
 Then the adjoint action of $\Carr(d+1)$ in the representation (\ref{Carrmatrix})
is
\beqa
\label{Ad}
\Ad(a^{-1})&&(\bomega,\bbeta,\bgamma,\alpha)=
\\[6pt]
&&(R^{-1}\bomega{}R, R^{-1}(\bomega\bb+\bbeta), R^{-1}(\bomega\bc+\bgamma), \alpha+\bb\cdot(\bomega\bc+\bgamma)-\bbeta\cdot\bgamma),
\nn
\eeqa
and  yields the coadjoint action, defined by $\Coad(a)\mu\equiv\mu\circ\Ad(a^{-1})$, as given by
 $\Coad(a)(\belle,\bk,\bp,m)=(\belle',\bk',\bp',m')$, where
\begin{eqnarray}
\label{ell}
\belle'
&=&
R\belle{}R^{-1}+(R\bk\,\barbb-\bb\,(R\bk)^T)+(R\bp\,\barbc-\bc\,(R\bp)^T)\nn\\ 
&&+\;m(\bc\,\barbb-\bb\,\barbc)\qquad
\\[4pt]
\label{g}
\bk'
&=&
R\bk+m\bc\\[4pt]
\label{p}
\bp'
&=&
R\bp-m\bb\\[4pt]
\label{m}
m'
&=&
m
\label{Coad}
\end{eqnarray}
showing that $m$ is a Casimir invariant with the dimension of \textit{mass} and $[m/\hbar]=
[s^{-1}]$ has the dimension of Carrollian frequency.

The Euclidean group is a subgroup of the Carroll group, and its Lie algebra $\mathfrak{e}(d)$ can be identified with (\ref{Z}) when $\beta=0$ and $\alpha=0$.
The dual space $\mathfrak{e}(d)^*$ consists of pairs $(\belle,\bp)$ and the coadjoint action if given by (\ref{g}) and (\ref{p}) with $\bk=0$ and $m=0$. If $d=3$, the coadjoint orbits are labelled by the invariants
$p=|\bp|$ and $\belle\cdot\bp$ where $\belle$ is viewed now as a $3$-vector. If $p\neq0$, then 
it is  interpreted as the \emph{color}, and 
$j={\belle\cdot\bp}/{p}$ as the \emph{spin.}
The coadjoint orbits are identified with $TS^2$ for all values  of $s$; for $j=0$ they consist of oriented straight lines in $\rE^3$ \cite{SSD,Chaundy,SpinOptics}.


The Carrollian  
massive case $m\neq0$ was studied in \cite{Carrollvs,Gomis,Ancille}.
Therefore we consider now the massless case $m=0$. Then we find three invariants, namely
\begin{equation}
p=\vert\bp\vert
\qquad
\&
\qquad
k=\vert\bk\vert
\qquad
\&
\qquad
w=\bk\cdot\bp.
\label{p2g2pg}
\end{equation}
By analogy with the Galilean/Euclidean cases, we call $p>0$ the  ``color''  \cite{SSD,SpinOptics}. 
These Carroll invariants have physical dimension $[p]=AL^{-1}$, $[k]=ML$, $[w]=MA$, where $[A]=[\hbar]$. 

In view of the form (\ref{ell})--(\ref{m}) of the coadjoint action of the Carroll group,
we choose first the origin of our massless coadjoint orbit as
\begin{equation}
\mu_0=(0,0,\bp_0,0)
\qquad
\hbox{with}
\qquad
\bp_0=p\,\bu_0,
\label{masslessmu0}
\end{equation}
where $\bu_0\in{}S^{d-1}\subset\bbR^d$ is a fixed direction. This massless orbit has  $k=w=0$, see (\ref{p2g2pg}).
The associated $1$-form (\ref{varpi}) on the Carroll group is 
$\varpi=p\,\bu_0\cdot{}R^{-1}d\bx$ and descends to the new evolution space 
\begin{equation}
V=\{(\bx,\bu,s)\in{}T\bbR^d\times\bbR\,\strut\big\vert\,\bu\cdot\bu=1\}
\quad
\hbox{as}
\quad
\varpi=p\,\delta_{AB}\,u^A dx^B,
\label{varpiTer}
\end{equation}
where $\bu=R\bu_0$. $V$ is the direct product of the  tangent bundle of the unit sphere in $d$ dimensions  carrying its canonical one-form with
(Carrollian) time.
Computing the characteristic foliation of  
\begin{equation}
\sigma=d\varpi=p\,\delta_{AB}\,du^A\wedge\,dx^B,
\label{dvarpiTer}
\end{equation}
in the $(2d)$-dimensional evolution space $(V,\sigma)$ we find it to be $2$-dimensional, tangent to the distribution $\ker\sigma$, and such that
\begin{equation}
Y
\in\ker\sigma\iff 
Y=\alpha\,u^A\frac{\partial}{\partial{}x^A}+\beta\frac{\partial}{\partial{}s}
\label{kersigmaTer}
\end{equation}
where $\alpha$ is a real Lagrange multiplier to enforce the constraint $\vert\bu\vert^2=1$, and ${\beta}\in\bbR$. 
The  space of motions $V/\ker\sigma$ now consists of straight lines (with $\bu=\const$), i.e., oriented geodesics of Euclidean space $\rE^d$. 
It is hence symplectomorphic to the cotangent bundle  
\beq
\cO_{\mu_0}\cong{}TS^{d-1}_p.
\eeq
of the $(d-1)$-sphere of radius $p$.

We note that the $1$-form $\varpi$ in  (\ref{varpiTer})  describes a ``Fermat'' particle whose motions are oriented line rays in Euclidean $3$-space (with no spin) studied in geometrical optics \cite{SpinOptics}.

The Fermat particle 
model can be obtained by a fixed-energy reduction of either a Galilean or a Poincar\'e-invariant particle with vanishing mass and spin
  \cite{SpinOptics}. This is explained by the fact that the space of motions, $TS^{d-1}$, endowed with the projection of the two-form (\ref{dvarpiTer}) is in fact
 a spin-zero coadjoint orbit of the \emph{Euclidean group}; but the latter is a subgroup of \emph{both} the Galilei and the Poincar\'e groups  \cite{SpinOptics}. 
 
 What we have just found shows that this same statement is valid for the Carroll group.

Lifting the Carroll vector field $X$ in (\ref{carralg}) 
to the evolution space $V$ as $X_V$, allows us to evaluate
$\varpi(X_V)$; see Eq. (\ref{muX}). These quantities are constant on the leaves of the characteristic foliation of $\sigma$, and therefore provide with the help of (\ref{muZ}) the components of the
 Carrollian momentum mapping $\mu=(\belle,\bk,\bp,m)$, namely
\beq
\barraynb{lllll}
\belle&=&\bp\,\bx^T-\bx\,\bp^T
&\qquad&\hbox{angular momentum,}
\\
\bk&=&0
&\qquad&\hbox{``centre of mass'',}
\\
\bp&=&p\,\bu
&\qquad&\hbox{linear momentum,}
\\
m&=&0
&\qquad&\hbox{mass.}
\earraynb
\label{carconsquant}
\eeq
In particular, $m=0$ is the conserved quantity associated to Carrollian time translations, $s\to s+f$. This follows immediately from that the variational form $\varpi$ in~(\ref{varpiTer}) has no component for $ds$.

\subsection{Conformal symmetries of massless Carrollian models}
 
We now show  that the \emph{zero-level}  conformal Carroll algebra, $\confcarr_0(d+1)\cong\mathfrak{e}(d)\ltimes{}C^\infty(\bbR^{d+1},\bbR$)  in (\ref{confcarr0}), is a symmetry algebra of the  Carroll massless particle model  above.
To see this, we lift  the $\confcarr_0(d+1)$
generators (\ref{confcarr0}) to $(V,\sigma)$,
as
\begin{eqnarray}
X_V&=&\left(\omega^A_B\,x^B+\gamma^A\right)\frac{\partial}{\partial{}x^A}+\omega^A_B\,u^B\,\frac{\partial}{\partial{}u^A}
+T(x^A,s)\,\frac{\partial}{\partial{}s}
\label{confcarr0Ter}
\end{eqnarray}
and we trivially check that
$ 
L_{X_V}\varpi=0
$ 
and 
hence $L_{X_V}\sigma=0$ for all $X\in\confcarr_0(d+1)$.  

A curious fact  is that although supertranslations extend the finite-dimensional symmetry (\ref{carralg})
to an infinite-dimensional one, they \emph{do  not contribute to the conserved quantities}. Again this follows  from the absence of any  $ds$-term in the ``varia\-tional'' $1$-form (\ref{varpiTer}). To put it in another way, we can say that \emph{any} (super or not) translation has $m=0$ as associated conserved quantity. This also explains 
why  Carrollian boosts $(\bx,s)\mapsto(\bx,s-\bb\cdot\bx)$ have a vanishing conjugate momentum,~$\bk=0$~:
they are just supertranslations of a particular form.

So far, we discussed ``Fermat photons'' with vanishing spin.
Spin can also be included. Restricting ourselves to $d=3$ spatial dimensions, it is sufficient to consider, instead of (\ref{masslessmu0}), the basepoint
\begin{equation}
\mu_0=(\belle_0,0,\bp_0,0)
\quad
\hbox{with}
\quad
\belle_0=j\be_1\quad\&\quad
\bp_0=p\,\be_1,
\label{masslessmu0spin}
\end{equation}
with $j\in\bbR$,
which yields the $1$-form on the (neutral) Carroll group
\beq
\varpi_{p,j}=p\bu\cdot d\bx-j\bv\cdot d\bw,
\label{spinvarpi}
\eeq
where $(\bu,\bv,\bw)\in\SO(3)$ and $\bx\in\bbR^3$.
The associated $2$-form  differs only from the spinless expression (\ref{dvarpiTer}) by an extra  term,
\beq
\sigma_{p,j}=d\varpi_{p,j}=p\,d\bu\wedge d\bx-j\,\surf,
\label{spinsigma}
\eeq
where  $\surf=\half\epsilon_{ABC}\,u^Adu^B\wedge{}du^C$, the canonical surface form of $S^2$ represents the spin $2$-form.

The characteristic foliation of (\ref{spinsigma}) is still $2$-dimensional; 
any tangent vector $Y\in\ker\sigma_{p,j}$ is again given by
\begin{equation}
Y=\alpha\,u^A\frac{\partial}{\partial{}x^A}+\beta\frac{\partial}{\partial{}s},
\label{kersigmaj}
\end{equation}
where $\alpha,\beta\in\bbR$.
We notice that the motions of these spinning particles are independent of the spin, $j$. The associated space of motions is again $TS^2$ with the new twisted symplectic form
\beq
\omega_{p,j}=\omega_{p,0}-j\,\surf.
\label{spinomega}
\eeq
Therefore $\CCarr_0(4)$ is  a symmetry group of spinning Carrollian photons also. 
The only effect of the new term in (\ref{spinvarpi}) (resp. in
(\ref{spinsigma})) is to change the angular momentum
[viewed as a $3$-vector] in (\ref{carconsquant}) to
\beq
\belle=\bx\times\bp+ j\bu,
\label{spinangmom}
\eeq
where the notations introduced in (\ref{masslessmu0spin}) and below (\ref{spinvarpi}) were used.

Our Carrollian ``photons'' are reminiscent of Galilean photons,
 studied before in \cite{SpinOptics,DHGalConf}. The latter are described
by the $2$-form
\begin{equation}
\sigma^{\Gal}_{p,j}=\sigma_{p,j}
-dE\wedge dt,
\label{Galsigma}
\end{equation}
defined on the Galilean extension of the evolution space with co\-ordinates $(\bu,\bx,E,t)$, obtained by adding $E$ and $t$, the energy and Galilean time, respectively, to the Euclidean evolution space. 
Their motions are hence instantaneous, i.e., ``move'' at $t=\const$ along straight lines in $\rE^3$.

\goodbreak

Spinless Galilean photons 
were shown \cite{DHGalConf} to carry an infinite-dimensional conformal Galilean symmetry,
 generated by the vector field
\beq
X=\Big(\omega_B^{\ A}(t)x^B+\eta^A(t)\Big)\frac{\ \p}{\p x^A}+T(t)\frac{\ \p}{\p t}\in\cgal_\infty(4)
\label{galconfphotsym}
\eeq
where $\bomega(t)\in\so(3)$, $\bfeta(t)\in\bbR^3$, and $T(t)\in\bbR$ depend arbitrarily on time~$t$, c.f. Eq.~(5.58) in \cite{DHGalConf}.
Galilean boosts and space translations correspond respectively to $\bfeta(t)=\bbeta t+\bgamma$, and time translations to $T(t)=\tau\in\bbR$. 
The time-independent vector fields (\ref{galconfphotsym}) generate  Souriau's  ``Aristotle group'', i.e., the Euclidean group extended with time-translations.

Viewed as an extension of the Euclidean model, the extra term $-dE\wedge{}dt$ in~(\ref{Galsigma}) is consistent with arbitrary time
dependence of the coefficients (since $t=t_0$ is now itself a constant of the motion), but  eliminates any position-dependence of the supertranslations
$T$. 

For the sake of comparison, we recall the conserved quantities found in \cite{DHGalConf}.

Define first a natural pairing between the infinite-dimensional Lie algebra 
$\cgal_\infty(4)$, see (\ref{galconfphotsym}), and its (formal) dual spanned by $\mu=(\belle(t),\bp(t),H(t))$ by
\beq
\mu\cdot{}X=\belle(t_0)\cdot\bomega(t_0)-\bp(t_0)\cdot\bfeta(t_0)+H(t_0)\,T(t_0)
\eeq for some $t_0$.
Then the associated  constants of the motion, 
\beq
\barraynb{lllll}
\belle(t_0)&=&\bx \times \bp + j\,\bu
&\qquad&\hbox{angular momentum,}
\\
\bp(t_0)&=&\bp
&\qquad&\hbox{linear momentum,}
\\
H(t_0)&=&E
&\qquad&\hbox{energy,}
\earraynb
\label{Galconsquant}
\eeq
are actually independent of the choice of $t_0$.

\section{Strings and particles}

Now we relate our particles to strings in the $(d+2)$-dimensional Bargmann space $(B,G,\xi)$, introduced in Section \ref{Unifchap}. This new approach will enable us to reveal distinguished, finite-dimensional, Lie subalgebras of the  conformal Galilei and Carroll Lie algebras discussed in Section \ref{SchCarr}.

We consider only the spinless case in flat Bargmann space, (\ref{Bstruct}). Let us start with the $2(d+2)$-dimensional manifold $T^*B$, with canonical coordinates 
$(p^1,\ldots,p^{d+2},x^1,\ldots,x^{d+2})$, endowed with  the Liouville $1$-form and symplectic $2$-form
\begin{equation}
\varpi=p_i\,dx^i
\qquad\hbox{and}\qquad
\Omega=dp_i\wedge{}dx^i,
\end{equation}
respectively.

\subsection{Massive systems}

We first consider the massive case, already discussed in Ref. \cite{DGH}. Let us consider the $2(d+1)$-dimensional submanifold $V$ of $T^*B$ defined by two constraints,
\begin{equation}
V=\left\{y=(p,x)\in{}T^*B\:\big\vert\;
p_i\xi^i=m,\;G^{ij}p_ip_j=0\right\}
\label{Vmass}
\end{equation}
with $m=\const\neq0$. In our usual coordinate system (\ref{Bstruct}), the first constraint says that
the momentum conjugate to the ``vertical'' variable, $s$, is $p_s=m$, and then the second relation identifies the momentum conjugate to $t$ as (minus) the kinetic energy of a non-relativistic particle of mass $m$  in $\bbR^d$, namely
$
p_t=-H=-\vert\bp\vert^2/({2m}).
$
Call~$\sigma$ the $2$-form on induced on $V$ by $\Omega$; we find
\begin{equation}
\sigma=\Omega\vert_V
= dp_A\wedge dx^A-dH\wedge dt.
\label{sigmam}
\end{equation}
Then $(V,\sigma)$ will be our \textit{evolution space}.

\goodbreak
 
The characteristic distribution of the closed $2$-form $\sigma$ is $2$-dimensional since
\begin{equation}
Y\in\ker\sigma
\quad
\hbox{iff}
\quad
Y=\alpha\,p^i\frac{\p}{\p x^i}+\beta\frac{\p}{\p{}s},
\label{chardistributionm}
\end{equation}
where $p^A=p_A$, $p_t=m$, $p_s=-H$ are constants of the motion, with
$\alpha,\beta\in\bbR$. By abuse of notation we will often identify $\ker\sigma$ given by (\ref{chardistributionm}) with its pointwise projection to Bargmann space~$B$.
In view of (\ref{chardistributionm}), \blue{its}
leaves project to Bargmann space as $2$-dimensional surfaces denoted by  $\Sigma$  which may be viewed as the \emph{world sheets of a string}.  The metric induced on the world sheet
has Lorentz signature $(+,-)$ since $G(Y,Y)=2m\,\alpha\beta$ at each $x\in\Sigma$.

The quotient space $U=V/\ker\sigma$ is symplectomorphic to $T^*\bbR^d$, i.e., to the \textit{space of  motions} of spinless \& massive Galilean particles.
In fact, we recover the ``Bargmannian'' description of such a particle, considered in
Refs. \cite{Eisenhart,DBKP,DGH}.

These strings $\Sigma$ project onto Newton-Cartan space-time $N$ as future-pointing time-like worldlines [in fact, as straight lines] since their tangent vectors are of the form
\begin{equation}
\pi_*Y
=
\alpha\left(p^A\frac{\p}{\p x^A}+ m\frac{\p}{\p t}\right)
\label{Galm}
\end{equation}
with $\bp=\const$ and $\alpha\in\bbR$.
In other words, they are \emph{worldlines of massive Galilean particles}.

The intersection of the above strings $\Sigma$ with the Carroll manifold $C\equiv B\vert_{t=0}$ yields again world lines.
 To see this we observe  that requiring that the distribution (\ref{chardistributionm})  also be tangent to $C$ implies $\alpha=0$ because~$m\neq0$. The corresponding distribution restricted to $C$ is therefore spanned by
\begin{equation} 
Y\vert_{C}=\beta\frac{\, \p }{\p s}
\label{Carrm}
\end{equation}
with  $\beta\in\bbR$. 

\goodbreak

The associated worldlines in $C$ are thus \emph{null} with respect to the  Carroll metric~\blue{$\rg$ in} (\ref{flatC}).  Moreover (\ref{Carrm}) shows that  free massive spin-$0$ Carrollian particles \emph{do not move} in Carroll absolute $d$-space, c.f. \cite{Gomis,Carrollvs,Ancille}.

\vskip3mm
Now we turn to symmetries. 


In the massive case $m\neq0$,
the constraints in (\ref{Vmass})  are clearly invariant under the conformal transformation
 generated by the vector fields $X$ on $B$ verifying
\begin{equation}
L_XG=\lambda\,G
\qquad
\&
\qquad
L_X\xi=0
\label{sch}
\end{equation}
for some function $\lambda$ of $B$. 
But this is precisely the infinitesimal version of (\ref{SchBarg}) 
and defines the \textit{Schr\"odinger Lie algebra}
(\ref{schbis}).

By construction, $\sch(d+1,1)$ is a symmetry Lie algebra of $(V,\sigma)$: if $X_V$ is the canonical lift of $X$ to $V\subset{}T^*B$, then 
\begin{equation}
L_{X_V}\sigma=0,
\qquad
\hbox{for all\ }
X\in\sch(d+1,1).
\label{schter}
\end{equation}

The projected vector fields 
  generate the centre-free Schr\"odinger Lie algebra of Newton-Cartan space-time $N$.

The vector fields in (\ref{schbis}) which also preserve  $C$, i.e., Carroll space $t=0$,  form
 the Lie algebra $\scarr(d+1)\cong\ce(d)\ltimes\bbR^{d+2}$ in (\ref{ccarr}), i.e.,
\begin{eqnarray*}
X=(\omega^A_B\,x^B+\gamma^A+\chi{}\,x^A)\frac{\partial}{\partial{}x^A}
+\big(\sigma-\beta_A\,x^A-\half\kappa\,x_Ax^A\big)\,\frac{\partial}{\partial{}s},
\end{eqnarray*}
which is thus 
a finite dimensional infinitesimal conformal symmetry of massive Carroll systems.  (Here $\ce(d)$ denotes the Lie  algebra of the conformal Euclidean group, 
i.e., the Euclidean group augmented with dilations.)

\subsection{Massless systems}\label{MasslessSec}

Now consider the \textit{new} $2(d+1)$-dimensional manifold
\begin{equation}
V=\left\{y=(p,x)\in{}T^*B\:\big\vert\;
p_i\xi^i=0,\;G^{ij}p_ip_j=p^2\right\}
\label{Vp}
\end{equation}
where $p=\const>0$ will be interpreted as the \textit{color} of massless ($m=0$) systems. We have hence
$p_s=0$ and $\vert\bp\vert^2=p^2$.
 The pull-back to $V$ of the canonical symplectic form on $T^*B$ is 
\begin{equation}
\sigma=d\varpi\vert_V=\Omega\vert_V
= dp_A\wedge dx^A+dp_t\wedge dt.
\label{sigmam0}
\end{equation}
The pair $(V,\sigma)$ is our new evolution space
for a massless particle with color $p=\vert\bp\vert$.
The kernel of $\sigma$ is $2$-dimensional, and pointwise spanned by the vectors
\beq
Y=\alpha\,p^A\frac{\p}{\p{}x^A}+\beta\frac{\p}{\p{}s}
\eeq
with $\alpha,\beta\in\bbR$; this can be read off from (\ref{chardistributionm}). 

The restriction of the leaves of $\ker\sigma$ to Carroll space are clearly $2$-dimensional manifolds $\Sigma$ which may be thought of as string worldsheets. 
The ``metric'' induced on these sheets from the Bargmann metric has signature $(+,0)$ since $G(Y,Y)=p^2\alpha^2$ at each $x\in\Sigma$; these strings are \textit{null strings}. 
In other words,
 our  ``particles'' with both vanishing spin and mass  are \textit{delocalized} in Carroll space-time; their ``history'' is a $2$-dimensional sheet, and not a curve.


Let us then investigate the conformal Carroll symmetries inherited from the Bargmann automorphisms. 

The constraints in (\ref{Vp}) describing massless systems are  invariant under the the transformations generated by the vector fields $X$ on $B$ satisfying
\begin{equation}
L_XG=0
\qquad
\&
\qquad
L_X\xi=\mu\,\xi
\label{isomAutB}
\end{equation}
for some function $\mu$ on $B$. 
The solutions $X$ of the system (\ref{isomAutB}) 
 are of the form
\begin{eqnarray}
X
&=&
\nonumber
\left(\omega^A_B\,x^B+\beta^At+\gamma^A\right)
\frac{\partial}{\partial{}x^A}
+
\left(\alpha{}t+\tau\right)\frac{\partial}{\partial{}t}\\
&&+\left(\sigma-\beta_A{}x^A-\alpha{}s\right)
\frac{\partial}{\partial{}s}\in\opt(d+1,1)
\label{opt}
\end{eqnarray}
where $\bomega\in\so(d)$, $\bbeta,\bgamma\in\bbR^d$, $\alpha,\tau,\sigma\in\bbR$; they are the generators of a Lie algebra that we propose to  call the \emph{optical Lie algebra} of the Bargmann space. 

By construction, $\opt(d+1,1)$ is a symmetry 
 of $(V,\sigma)$: if $X_V$ is the canonical lift of $X$ to $V\subset{}T^*B$, then 
\begin{equation}
L_{X_V}\sigma=0,
\qquad
\hbox{for all\ }
X\in\opt(d+1,1).
\label{optbis}
\end{equation}
The associated moment maps, i.e., conserved quantities are those of the Carroll group listed in  (\ref{carconsquant}). Note, once again, that both the ``$t$'' and  ``$s$'' super\-translations have vanishing contribution to the conserved quantities. This follows at once from the particular form  of $\sigma$ in (\ref{sigmam0}).

Projecting the vector fields (\ref{opt})
onto Newton-Cartan space-time yields
\beq
\pi_*X
=
\left(\omega^A_B\,x^B+\beta^At+\gamma^A\right)
\frac{\partial}{\partial{}x^A}
+
\left(\alpha{}t+\tau\right)\frac{\partial}{\partial{}t}
\label{NCopt}
\eeq 
which generate the \emph{centre-free optical Lie algebra}, i.e., the Galilei Lie algebra augmented with time dilations, found earlier as a symmetry of the Chaplygin gas with viscosity \cite{HHPLA}.

Finally, the vector fields (\ref{opt}) preserving the  Carroll space $C$ viewed as the $t=0$ slice of $B$, are of the form
\begin{equation}
X\vert_{C}
=
\left(\omega^A_B\,x^B+\gamma^A\right)
\frac{\partial}{\partial{}x^A}
+\left(\sigma-\bbeta\cdot\bx-\alpha{}s\right)
\frac{\partial}{\partial{}s} 
\label{g0}
\end{equation}
and span, as anticipated, a Lie subalgebra of the conformal Carroll Lie algebra of level $k=0$. It is  isomorphic to the semi-direct product 
of the Euclidean Lie algebra in dimension $d$, and of the space of polynomials of degree $1$ on Carroll space~$\cC\cong\bbR^{d+1}$,
\begin{equation}
\se(d)\ltimes\mathrm{Pol}_1(\cC)\subset\ccarr_0(d+1).
\label{gbis}
\end{equation}
It is worth noting that the symmetry (\ref{g0})
actually extends to the full $\ccarr_0(d+1)$ algebra
(\ref{galconfphotsym}) since the supertranslations
do not contribute.

\subsection{Schild strings}\label{SchildSec}
 
As noted in section \ref{MasslessSec},  for representations with vanishing mass and spin  the space of motions (i.e., the set of leaves of $\ker\sigma$) consists two-dimensional world sheets moving in Bargmann spacetime and such that  
 the metric induced from the  Bargmann metric is degenerate and has  rank $1$. The dynamics
of such strings, which are often referred to as ``null strings'' or ``tensionless strings'', was first discussed 
by Schild in a  posthumous  paper in 1977 \cite{Schild}.

 With the development of modern
string theory Schild's work has been extensively revisited
in the theoretical physics literature
from  a variety of perspectives, both classical and quantum.
Just as null geodesics may be thought of as 
a limiting case of timelike geodesics, Schild
strings may be thought of as limiting case 
of Nambu-Goto strings. The analogy goes deeper,
in that just as the usual proper time action
whose variation yields the equations of motion
for  timelike geodesics is not suitable
for null geodesics, the analogous Nambu-Goto action
for timelike strings is not suitable for null strings.
Schild was able to provide a convenient replacement.
Below we provide a summary  of the formalism and compare our work with
some existing treatments of the conformal symmetries
with those in the literature.


Here we briefly summarize Schild's 1977  formalism  \cite{Schild} for null
strings which differs somewhat from more recent treatments. 
Schild considers a $2$-dimensional 
submanifold $\Sigma$  
\beq
x^i = x^i(u^a),\qquad i = 1,2,\dots ,n, \qquad a=1,2, 
\eeq
of an $n$-dimensional  Lorentzian $(M,\rg)$ spacetime which satisfy
the Schild variational equations. These obtained by considering the bi-vector 
\beq
\sigma^{ij} = - \sigma^{ij}= 
 \half \epsilon^{ab}\frac{\p x^i}{\p u^a}
 \frac{\p x^j}{\p u^b}, \qquad \epsilon^{ab}=- \epsilon^{ba},\qquad  \epsilon
^{12}= 1,
\eeq
and its magnitude squared 
\beq
\sigma^2 =  \half \rg_{i k}\,\rg_{j l}\,\sigma^{ij} \sigma ^{kl}
\eeq
and demanding that
\beq
 \delta\int_\Sigma \half\sigma^2  du^1du^2=0.
\eeq
Note that from the point of view of the world sheet $\Sigma$, the quantity
$\sigma^2 $ is a scalar density of weight $2$ and hence 
\emph{Schild's  action is not invariant
under all diffeomorphism of the world sheet $\Sigma$}.
In more recent treatments compensating fields are included
to restore full diffeomorphism invariance.
This means that the world sheet symmetries 
may differ. However it does not affect the spacetime symmetries.

The Euler-Lagrange equations imply that
\beq
\sigma^2 = \const, 
\eeq
and we obtain null strings by taking the constant to vanish.
One may check that then the induced metric  
\beq
\gamma_{ab}= \rg_{ij} \frac{\p x^i}{\p u^a}  \frac{\p x^j}{\p u^b}  
\eeq 
is   tangent to the local light cone and is therefore
degenerate, i.e. has signature~$(+,0)$. The null direction is  
the kernel of $\sigma_{ij}$  and defines 
null vector field on $\Sigma$ whose integral curves, called null generators, 
fibre $\Sigma$. 

Schild's equations of motion imply that these null curves are
null geodesics of the ambient spacetime manifold $(M,\rg)$.    
Thus the \emph{general solution of Schild's equations of motion for null strings
is  obtained by taking an arbitrary spacelike curve 
$\gamma$ in $(M,\rg)$ and a one-parameter family of null vectors along it.
One then pushes $\gamma$ along the null geodesics of $(M,\rg)$
with these initial tangent vectors.}
If $(M,\rg)$ is flat then these null geodesics are 
straight lines with a null tangent vector. In the case of Carroll strings
lifted to Bargmann space 
one takes $\gamma$ to be a straight line, i.e., a geodesic
in ${\rE}^n$ and the null geodesics to be straight lines
with ${\bf x} \in \gamma $ and $t$ both constant. The coordinate $s$ then serves
as an affine parameter along the null generators.


As written by Kar \cite{Kar}, the partially gauge-fixed Schild equations are
\beq
\ddot{x}^i +\Gamma^i_{jk} {\dot x}^j {\dot x}^k=0, 
\qquad g_{ij}{\dot x}^i{\dot x}^j =0, \qquad  \rg_{ij}{\dot x}^i{x^\prime}^j =0, 
\eeq
where ${\dot x}^i =\p_\tau x^i $ and ${x^\prime}^i = \p_\sigma x^i$. 
Thus at $\sigma=u^1$ constant we have null geodesics with  affine 
parameter $\tau=u^2$ and we have the induced metric 
\beq
ds^2 = N^2(\sigma) d\sigma ^2.
\eeq

Kar claims that these equations are invariant under
the remaining gauge freedom
\beq
\tau \rightarrow \tau(\tau,\sigma), \qquad \sigma \rightarrow \sigma(\sigma),
\eeq 
which are generated by
\beq
A(\tau,\sigma) \p_\tau + G(\sigma) \p_\sigma.
\eeq
This is a \emph{Newman-Unti group}  in $1+1$ dimensions, discussed in  Sec. \ref{NUsec}.

By contrast Bagchi \cite{BagSchild}, following earlier work \cite{Isberg},
claims invariance under a group generated
by
\beq
(\tau f^\prime(\sigma)+g(\sigma))\p_\sigma+g(\sigma) \p_\tau.
\eeq
This preserves the vector half-density
\beq
V^a= \delta^a_\tau.
\eeq
His CGA  is in turn the same as 
$\ccarr_2(1+1)$, as pointed out in Sec. \ref{Examples},
since we are in $1+1$ dimensions.

\section{Conclusion}

In this paper we introduced systematically a family of infinite dimensional conformal Carroll groups $\CCarr_k(d+1)$ labelled by an integer $\cN=0,1,\dots,$  which acts upon $(d+1)$-dimensional Carrollian space-time. When $d=1$, they are the same as the Conformal Galilei groups $\CGal_{\cN}$. These generalize in a natural way to more general Carroll manifolds and in particular give rise to the Bondi-Metzner-Sachs (BMS) groups and their Newman-Unti (NU) generalizations. The case $\cN=0$ is shown to apply to massless Carrollian systems. The latter  are in fact strings, which lift to Schild's null strings
in $(d+2)$-dimensional Bargmann space.


\begin{acknowledgments} 
G.W.G would like to thank Quim Gomis for many illuminating  conversations
 on the Carroll group, and  {\it KITP} in  Santa Barbara for its hospitality during its
2012 \textit{Bits and Branes workshop} discussions 
at which stimulated the present investigation.
He is grateful also to the 
{\it Laboratoire de Math\'ematiques et de Physique Th\'eorique de l'Universit\'e de Tours}  for hospitality, and the  {\it R\'egion Centre} for a 
\emph{``Le Studium''} research professor\-ship.      
\end{acknowledgments}



\end{document}